\documentclass{ws-procs9x6}

\newcommand{\be}{\begin{equation}}
\newcommand{\ee}{\end{equation}}
\newcommand{\bea}{\begin{eqnarray}}
\newcommand{\eea}{\end{eqnarray}}

\begin{document}

\title{Spontaneous Lorentz Violation and Baryogenesis}

\author{Jing\ Shu}

\address{Enrico Fermi Institute, Deptartment of Physics, \\ 
and Kavli Institute for Cosmological Physics, \\
University of Chicago, 5640 S. Ellis Avenue, Chicago, IL 60637, USA\\
HEPDivision, ArgonneNational Laboratory, \\9700CassAve., Argonne, IL60439, USA \\
E-mail: jshu@theory.uchicago.edu}  

\maketitle

\abstracts{
In the presence of background fields that spontaneously violate
Lorentz invariance, a matter-antimatter asymmetry can be generated
even in thermal equilibrium.  In this paper we systematically
investigate models of this type, showing that either high-energy
or electroweak versions of baryogenesis are possible, depending on
the dynamics of the Lorentz-violating fields. We identify two scenarios of
interest: baryogenesis from a weak-scale pseudo-Nambu-Goldstone
boson with intermediate-scale baryon-number violation, and
sphaleron-induced baryogenesis driven by a constant-magnitude
vector with a late-time phase transition.}

\section{Introduction}

The observed universe manifests a pronounced asymmetry between
the number density of baryons $n_b$ and antibaryons $n_{\bar b}$. 
However, the origin
of the baryon number asymmetry remains a major puzzle for cosmology
and particle physics. In a classic work, Sakharov argued that three conditions are
necessary to dynamically generate a baryon asymmetry in an
initially baryon-symmetric universe: (1) baryon number
non-conserving interactions; (2) C and CP violation; (3) departure
from thermal equilibrium. In deriving these
conditions, the assumption is made that CPT is conserved. 
If Lorentz invariance is violated, then CPT is also violated, one can generate baryons in thermal equilibrium. This idea is first 
implemented in the context of  ``spontaneous
baryogenesis''\cite{CK88} scenario and has subsequently been
elaborated upon in various ways \cite{BCKP96,Trodden,XMZ,Yamag,Chiba:2003vp,Alberghi:2003ws,DKKMS04}.

In most of the previous studies, the effects of sphaleron transitions are not discussed or mistakenly believed to wash out the baryon asymmetry if $B-L$ is zero. The role of sphaleron transitions in thermal equilibrium is to adjust different particle density distributions in a way that preserves $B-L$ to minimize the total free energy. In the presence of a Lorentz-violating background, the one particle free energy is modified and final net baryon number density should be quite different from the one without Lorentz-violating background. In this situation, sphaleron transitions will generate a nonzero $B+L$. Because of that, we  reconsider previously studied models and construct new models\cite{Carroll:2005dj}. We find the final net baryon number density largely depends on how the effective chemical potential $g a_0$ evolves with time. This provide us a general rule to categorize different models of baryogenesis via spontaneous Lorentz violation and understand them in a unified picture. We also identify two scenarios of potential interest.  One is the case
of a simple constant-magnitude timelike vector field coupled to
$J^\mu_{B+L}$ where appropriate baryon asymmetry could be generated by electroweak sphalerons alone and the other is that of a
derivatively-coupled pseudo-Nambu-Goldstone boson with a weak scale mass and ($B-L$)-violating interactions are freeze out at Majorana neutrino mass scale of order $10^{10}$~GeV. 

\section{Baryogenesis in the presence of Lorentz violation}

We consider the theory of a vector field $A_\mu$ with a nonzero
vacuum expectation value (vev), coupled to a current $J^\mu$ in the matter fields which corresponds to some continuous global symmetry. The vector field gets a condensate $A_\mu = (a_0, 0,0,0)$ and that makes the interaction term $\mathcal{L}_{int} = g A_{\mu} J^{\mu} \rightarrow -g a_0 Q$, where $Q$ is the conserved charge. Such an interaction term now acts like a ``chemical potential" $\mu^0_{b} = g a_{0}$ for the matter fields, which splits the free energy of particle and anti-particle. Because of this effect,
there will be a non-zero baryon number
density generated by baryon-number violating interactions in thermal equilibrium $ n_{B} = n_{b} - n_{\bar{b}} ={g_{b}T^{3}} \Big[
\pi^{2} \frac{\mu_{b}^0}{T} + \Big(\frac{\mu_{b}^0}{T} \Big) ^{3}
\Big] / 6 \pi^{2} \simeq g_{b}\mu_{b}^0 T^{2}/{6} \sim \mu_{b}^{0} T^{2},
$
where $g_{b}$ counts the internal degrees of freedom of the baryons.
When the $B$-violating interactions mentioned above become
ineffective ($\Gamma \leq H$), we get the final baryon asymmetry
\begin{equation}\label{Eq1-2}
{n_{B} \over s} \sim \frac{\mu_b^0}{g_{\ast s}T_{F}} = \frac{g
a_0}{g_{\ast s}T_{F}}
 \, ,
\end{equation}
with the entropy density
$s = (2\pi^{2}/45) g_{\ast s}
T^{3}$ 
, where $T_{F}$ is the temperature at which the baryon number
production is frozen out.

With such a spontaneous Lorentz violation background,  the energy dispersion relation is modified to $E= \sqrt{K_i^2 + m^2}
\pm \mu^{0}$, where $K_i$ is the momentum of the fermion and $\pm$ are for fermion and anti-fermion respectively. The net baryon number density now becomes $B^{(\mu)} = - \frac{2N}{13} T^2 (3 \mu^{0} + \frac{1}{N} \sum_{i=1}^{N}
\mu_{i}^{0})$, 
where the parameters $\mu$ and $\mu_{i}$ are the chemical
potentials of the quarks and the $i$th lepton, respectively. If there is leptonic flavor violation in thermal equilibrium, one can write it in terms of $\mu_{L}^{0} \equiv
\displaystyle \frac{1}{N} \sum_{i=1}^{N} \mu_{i}^{0}$ and $\mu_{B}^{0} = 3\mu^{0}$. The fact that $B \propto (\mu^{0}_{B} + \mu^{0}_{L}) =
2 \mu^{0}_{B+L}$ tells us a
nonzero net baryon number density can be spontaneously
generated through sphaleron transitions in thermal equilibrium
in the presence of a
nonzero time-like vector background coupled to $J_{B+L}$
current.

As we know sphaleron transitions connect
baryon and lepton number, we need to consider both the
baryon number current and lepton number current that couple to the
background field. It is convenient to rewrite $B$ and $L$ currents in terms of the
$B+L$ and $B-L$ currents. From Eq. (\ref{Eq1-2}), we know that
\begin{equation}\label{E1-4}
\frac{n_{B-L}}{s} = \frac{\mu_{-}^0(T_{-})}{g_{* s} T_{-}} =
\frac{g_{-} a_0(T_{-})}{g_{* s} T_{-}}\ ,
\qquad \frac{n_{B+L}}{s} = \frac{\mu_{+}^0(T_{+})}{g_{* s}
T_{+}}= \frac{g_{+} a_0(T_{+})}{g_{* s} T_{+}}
 \, ,
\end{equation}
where $T_{-}$ and $T_{+}$ are the lowest freeze-out temperature
for any interactions that could violate $B-L$ and $B+L$,
respectively. $T_{+}$ is the sphaleron freeze-out temperature
which is roughly 150~GeV. $T_{-}$ ranges from TeV to GUT scale and is very model dependent. Notice that $T_{-} \ll T_{+}$, so whether $n_{B+L} \ll n_{B-L}$ or $n_{B+L}
\gg n_{B-L}$ will only depend on whether $\mu^0(T)/T$ is an
increasing or decreasing function with respect to $1/T$.
The net baryon number $n_{B} = (n_{B+L} + n_{B-L})/2$, so we know
that $n_{B}$ is of the same order as max$\{n_{B+L}$, $n_{B-L}\}$.
{}From Eq.~(\ref{E1-4}), we get
\begin{eqnarray}\label{E1-5}
  \frac{n_B}{s} &=& \left\{
  \begin{array}{l}
  \displaystyle \frac{n_{B+L}}{2s} \sim \frac{g_{+}
  a_0(T_{+})}{g_{* s} T_{+}}
  \, ~~ \textrm{if~}
  \frac{|\mu^0(T)|}{T} ~~
  \textrm{increases as a function of $1/T$,} \\
  \displaystyle \frac{n_{B-L}}{2s} \sim \frac{g_{-}
  a_0(T_{-})}{g_{* s} T_{-}}  \, ~~  \textrm{if~}
  \frac{|\mu^0(T)|}{T} ~~
  \textrm{decreases as a function of $1/T$.}
  \end{array}\right.
\end{eqnarray}

\section{Present-day constraints on Lorentz violation}
\label{constraints}

In principle, there are no real experimental constraints as baryogenesis happens at the early time while all experiments are at present. Nevertheless, the highly constrained experimental bounds today suggest that the spontaneous Lorentz-violating background undergoes a phase transition if it doesn't decay away or roll to an extremely small value.  

The direct constraints between baryon number current and Lorentz violating background field are coming from neutral meson mixing. Only the difference $\Delta a_\mu$ between the
corresponding two $a_\mu$ coefficients is observable. The experimental
constraint comes from the parameter $
  \Delta a_{\mu} = r_{q_1}
  a_{\mu}^{q_1} - r_{q_2} a_{\mu}^{q_2}\,,
$
where $a_{\mu}^{q_1}$,
$a_{\mu}^{q_2}$ are Lorentz-violating coupling constants for the
two valence quarks in the meson, and where the factors $r_{q_1}$
and $r_{q_2}$ allow for quark-binding or other normalization
effects. Experiments studying neutral $K$-mesons have
constrained two combinations of $\Delta
a$ for $d$ and $s$ quarks, with bounds
in the Sun-centered frame of approximately
\be
  |\Delta a_{0}| \leq 10^{-20} {\rm ~GeV}
  \label{deltaa0}
\ee
by the KTeV Collaboration at Fermilab \cite{CPT2,KTeV}. Other
experiments with $D$ mesons have constrained two combinations of
$\Delta a$ for the $u$ and $c$ quarks at about $10^{-15}$ GeV (FOCUS
Collaboration, Fermilab) \cite{CPT2,FOCUS}. There are even more constrained results from axial vector current and astrophysics, but the couplings in those constraints are essential to generate baryons. 

\section{Sources of Lorentz violation}

We first consider that Lorentz-violating vector field $A_\mu$
has a constant expectation value in the vacuum ($a_0 =$~constant). Our discussion could also be generalized to a ghost field and high rank-tensor condensate\cite{BCKP96}. The vector condensation is achieved through a Mexican-hat potential\footnote{Notice we use the minus metric here.} $V(A_{\mu}) = \frac{\mu^2}{2}A_{\mu}A^\mu + \frac{\lambda}{4}(A_{\mu}A^\mu)^2$. At the classical level, the timelike component for a vector field with minimal ground state energy is given by ${a_{0}}^2 = {\mu^2\over \lambda}$. Since $\mu^0/T = ga_0/T$ is increasing
as the universe expands (and $T$ decreases), the relevant freeze-out
temperature is thus $T_{+} = 150$~GeV, due to sphaleron transitions.
{}From Eq.~(\ref{Eq1-2}) we obtain
\begin{equation}\label{E3-1}
  {n_{B}\over s} \sim \frac{ga_{0}}{g_{\ast s}T_{+}}
  \sim ga_0 (10^4 {\rm ~GeV})^{-1} \approx 10^{-10} 
 \, .
\end{equation}
Such a field condensate would seem to violate the
experimental constraints discussed in Section~\ref{constraints}.
One way to accomodate the experimental limits is to imagine a phase transition for the $A_\mu$ field
itself.  We can replace the
coefficient of the mass term $\mu^{2}$ with $({\mu'}^{2}- \alpha|\Phi|^2)$ in the potential,
where $\Phi$ is the Higgs doublet.
At high temperatures, the Higgs expectation value $\langle\Phi\rangle$
vanishes, and we get a non-zero vacuum expectation value of the vector
background field. At late times, $|\Phi|^2 = v^2$. If $\mu'^2 -
\alpha v^2$ is negative, we get a zero vev for vector background field.
So the Lorentz symmetry is restored. 

Another possibility is that the Lorentz-violating background is time-dependent.  A simple way to achieve this is to image $A_\mu$ is the gradient of a slowly-rolling scalar field $\phi$. The chemical potential term $a_{0}$ is then given by $\dot{\phi}/
f$. When one applies this to a quintessence field which has ``tracking'' behavior, $\dot{\phi} \propto \sqrt{V(\phi)} \propto \sqrt{\rho_{\rm back}} \propto T^2$. We can see that the freeze-out temperature is determined by $T_{-}$, which is very model dependent. Similarly, one can consider the interaction between the
derivative of the Ricci scalar curvature $\mathcal{R}$ and the
baryon number current $J^{\mu}$ from the effective theory of
gravity\cite{DKKMS04}
$
\frac{1}{M^{2}_{*}} \int d^{4}x \sqrt{-g} (\partial_{\mu}
\mathcal{R} ) J^{\mu} 
$.
The net baryon number density obtained is proportional to an even
higher positive power of freeze-out temperature $T_{F}$, since
$\dot{\mathcal{R}} \propto \dot{\rho}$, where $\rho$ is the total
energy density.

Finally we consider the Lorentz-violating background arising as the gradient of
a pseudo-Nambu-Goldstone boson (PNGB) $\phi = f \theta$, where $f$ is the scale of spontaneous symmetry breaking. A PNGB remains overdamped in its
potential until its mass becomes comparable to the
Hubble parameter, at which time it will roll to its minimum and
begin to oscillate and decay like a massive particle. If we want to generate a baryon asymmetry of the right
amplitude, then from
${n_{B}}/{s} \sim
\dot{\phi} / f g_{\ast s} T_{F} = 10^{-10}
$
and $g_{\ast s}\sim 100$ we require
$
\dot{\phi} = 10^{-8} f T_{F}
$. The PNGB obeys the equation of motion
$\ddot\phi + 3H\dot\phi + \frac{dV}{d\phi}=0$ and we drop the term $\ddot\phi$ as we are interested in the rolling phase.  For typical
values $\phi \sim f$, we have
$\dot\phi \sim H^{-1}\frac{dV}{d\phi} \sim
 {m^2 f M_{pl}}/{T^2}$. Thus, to achieve successful
baryogenesis requires that the freeze-out temperature satisfy $
  \frac{T_F^3}{m^2} \sim 10^8 M_{pl}\,.
$
The ideal circumstance would be if freeze-out occured when the field had just begun to roll substantially, but not yet begun to oscillate.  This
corresponds to $H\sim m$, which implies $T_F^2 \sim m M_{pl}\,.$
Comparing to the expression of $\dot{\phi}$ shows that PNGB baryogenesis works
if the freeze-out temperature is at an intermediate scale
$T_F \sim 10^{-8} M_{pl} \sim 10^{10}~{\rm GeV}$
and the PNGB mass is
$m \sim {T_F^2}/{M_{pl}}
  \sim 100~{\rm GeV}$.

\section{Conclusion}

We have investigated the possible origin of the observed baryon
asymmetry in the presence of a coupling between a
Lorentz-violating vector field and the baryon current, and especially reconsidered the effects of sphaleron transitions. If $a_{0}/T$ is increasing with time, then the
final net baryon number density is determined by the freeze-out
temperature $T_{+} \approx 150 $~GeV. For the opposite case, the
final net baryon number density is determined by the freeze-out
temperature $T_{-}$ which is model-dependent. 
Most previous works \cite{BCKP96,Trodden,XMZ,Yamag,DKKMS04} consider a slow rolling scalar field as the spontaneous Lorentz-violating background, and the
absolute value of the effective chemical potential is decreasing with time,
so the right net baryon number density generated depends on a high freeze-out temperature $T_{-}$. 
In order to obtain the right net baryon number density, the coupling times the time component of the
background field $g a_{0}$ should be not too small, and experimental constraints at present suggest that we need some dynamical mechanism to
decrease $a_{0}$. We first consider a constant Lorentz-violating background case and sphaleron transitions will be the main source to generate the baryon asymmetry. We can imagine
that a phase
transition occurs in between freeze-out and today. Our investigation of the PNGB scenario reveals that the most
natural implementation of this idea requires PNGB's with
weak-scale masses (100~GeV) and $(B-L)$-violating interactions
that freeze out at an intermediate scale of around
$10^{10}$~GeV naturally from the decay of Majorana neutrinos.  We therefore
consider this scenario to be quite promising.

\end{document}